\documentclass[article,amsmath,amssymb]{revtex4-2}
\usepackage{graphicx}
\usepackage{dcolumn}
\usepackage[colorlinks]{hyperref}
\usepackage{bm}

\begin{document}

\title{Classical versus quantum intensity-field correlations of scattered light from extended cold atomic clouds}

\author{N. Piovella}

\affiliation{Dipartimento di Fisica "Aldo Pontremoli", Universit\`{a} degli Studi di Milano, Via Celoria 16, I-20133 Milano, Italy}

\date{\today}

\begin{abstract}
We calculate the intensity-field correlations in the
light scattered by $N$ cold atoms driven by a quasi-resonant
laser field. Fundamental differences occur if the atomic state is
an entangled single-excitation state or a coherent factorized state. We provide analytic expressions for the two-time field and intensity correlation functions for the timed Dicke state and the quasi-Bloch state. The comparison with multi-atom simulations shows good agreement between numerical and analytic solutions.
\end{abstract}

 \maketitle

\section{Introduction}

Statistical properties of the radiation scattered by $N$ identical two-level atoms are usually studied by means of the single-time second-order normalized correlation function $g^{(2)}$ \cite{Cohen1998}. When the system is in an excited state, $g^{(2)}$ roughly equals two. This value is typical of the Hanbury Brown and Twiss effect for thermal or pseudothermal Gaussian fields \cite{Hanbury1956}. When the system decays from the excited state to the superradiant state \cite{Dicke1954,Eberly1970}, the behavior of $g^{(2)}$ is essentially classical, since it decreases from 2 to nearly 1, which is the value from a coherent state. However, as the system further decays to the ground state, the intensity fluctuation function, calculated in a model where the field is quantized, is quite different from the classical one \cite{Tallet1972}. Similar results are expected also for subradiance, the odd-twin of superradiance, in which the excited atoms stay trapped or decay slower to the ground state. It has been shown that subradiance can be considered to be a purely nonlocal, nonclassical phenomenon displayed by quantum sources \cite{Bhatti2018}.

More recently, a renewed interest in superradiance and subradiance has increased in the context of the cooperative light scattering in cold atomic samples \cite{Bienaime2010,Guerin2016}. These studies stem from the seminal work by Scully \textit{et al.} \cite{Scully2006}, which described the single-photon superradiance from $N$ two-level atoms prepared by the absorption of one photon of wave vector $\mathbf{k}_0$  \cite{Svidzinsky2008,Scully2009,Eberly2006}. It has been shown that the photon is spontaneously emitted in the same direction of the incident photon with a cooperative decay rate proportional to $N$ and inversely proportional to the size of the atomic cloud. These studies considered the decay of atoms prepared in the ‘timed Dicke state’:
\begin{equation}
|+\rangle_{\mathbf{k}_0}=\frac{1}{\sqrt{N}}\sum_{j=1}^N e^{i\mathbf{k}_0\cdot \mathbf{r}_j}|g_1, g_2,\dots, e_j, \dots,g_N\rangle
\label{TDS}
\end{equation}
where $|g_1, g_2,\dots, e_j, \dots,g_N\rangle$ is a Fock state in which the $j$th atom
is prepared in the excited state $|e_j\rangle$ and all the other atoms are in the ground state $|g_j\rangle$, and $\mathbf{r}_j$ is 
the position of the $j$th-atom. A natural platform for studying this kind of superradiance is provided by the cooperative scattering, in which the atoms cooperate to scatter the photons from an incident laser beam with wave number $\mathbf{k}_0$ and frequency $\omega_0=ck_0$ close to the atomic resonance, leading to a directional emission. This phenomenon is due to the synchronization of the atomic dipoles with the laser.
By different experiments and comparing different models \cite{Chabe2014,Bachelard2016} it has been shown that in the linear optics approximation, i.e. with a weak excitation of the atoms, the behavior is almost classical and can be described by considering the atoms as classical dipoles. Quantum effects detectable by measuring the scattered intensity or the cooperative radiation force exerted on the atoms could be observed only with a strong pump such that saturation effects become important \cite{Pucci2017}, or by investigating the fluctuations of the system. The objective of this work is to give evidence of the quantum or classical effects observable in the linear optics approximation, by evaluating the two-time correlation functions $g^{(1)}(t,\tau)$ and $g^{(2)}(t,\tau)$ for the photons scattered by cold two-level atoms in different quantum states, i.e., (a) the entangled single-excitation state, (b) the timed-Dicke state \cite{Scully2006}, (c) the product (or coherent) state, and (d) what we call a quasi-Bloch \cite{Friedberg2007} or Eberly's \cite{Eberly2006} state. Whereas $g^{(1)}(t,\tau)$ is the same for all these states, the intensity-correlations function $g^{(2)}(t,\tau)$ exhibits  fundamental differences related to the quantum or classical description of the atomic ensemble.

\section{The model}\label{Hamiltonian}
Our system  consists of a gas of $N$ two-level atoms (with random, fixed positions $\mathbf{r}_j$, lower and upper states $|g_j\rangle$ and
$|e_j\rangle$ with $j=1,\dots,N$, and transition frequency $\omega_a$ with linewidth $\Gamma=d^2\omega_a^3/2\pi\hbar\epsilon_0 c^3$, where
$d$ is the electric dipole matrix element), driven by a uniform resonant radiation beam  with wave vector $\mathbf{k}_0=k_0\mathbf{\hat e}_z$, frequency $\omega_0=\omega_a+\Delta_0$ and electric field $E_0$. The atom-field interaction Hamiltonian in the rotating-wave approximation (RWA) is
\begin{equation}\label{H}
\hat H=\hat H_0+\hat H_1
\end{equation}
where
\begin{eqnarray}
    \hat{H}_0 &=&\hbar\sum_{j=1}^N\left\{
    -\frac{\Delta_0}{2}\hat\sigma_{3j}+\frac{\Omega_0}{2}
    \left(
    \hat\sigma_je^{-i\mathbf{k}_0\cdot\mathbf{r}_j}
    +{\hat\sigma_j}^{\dagger}e^{i\mathbf{k}_0\cdot\mathbf{r}_j}\right)\right\}\label{H0}\nonumber\\
    \hat{H}_1 &=&\hbar\sum_{j=1}^N\sum_{\mathbf{k}}g_k\left[\hat{a}_{\mathbf{k}}^\dagger\hat\sigma_j
    e^{i(\omega_k-\omega_0)t-i\mathbf{k}\cdot\mathbf{r}_j}
        +{\hat\sigma_j}^{\dagger}\hat{a}_{\mathbf{k}}
    e^{-i(\omega_k-\omega_0)t+i\mathbf{k}\cdot\mathbf{r}_j}\right]\label{H1}.
\end{eqnarray}
Here $\Omega_0=dE_0/\hbar$ is the pump Rabi frequency,
$\hat{a}_{\mathbf{k}}\exp(-i\omega_kt)$ is the photon annihilation operator in the interaction picture, with
wavenumber $\mathbf{k}$ and frequency $\omega_k=ck$, $g_k
=d\sqrt{\omega_k/(2\hbar\epsilon_0V_{ph})}$, $V_{ph}$ is the photon
volume, $\hat\sigma_j=\exp(i\Delta_0 t)|g_j\rangle\langle e_j|$
and $\hat\sigma_{3j}=|e_j\rangle\langle e_j|-|g_j\rangle\langle
g_j|$. We write the Heisenberg equations of the atomic and field operators as
\begin{eqnarray}
  \frac{d\hat\sigma_{j}}{dt} &=& \frac{1}{i\hbar}[\hat\sigma_j,\hat H]=
  i\Delta_0\hat\sigma_j+\frac{i\Omega_0}{2}\hat\sigma_{3j}e^{i\mathbf{k}_0\cdot\mathbf{r}_j} +i\sum_{\mathbf{k}}g_k
  \hat\sigma_{3j}\hat a_{\mathbf{k}}e^{-i(\omega_k-\omega_0)t+i\mathbf{k}\cdot \mathbf{r}_j}\label{s1bis}\\
  \frac{d\hat\sigma_{3j}}{dt} &=& \frac{1}{i\hbar}[\hat\sigma_{3j},\hat H]=
   i\Omega_0 \hat\sigma_{j}e^{-i\mathbf{k}_0\cdot\mathbf{r}_j}+2i\sum_{\mathbf{k}}g_k
   \hat a_{\mathbf{k}}^\dagger\sigma_{j} e^{i(\omega_k-\omega_0)t-i\mathbf{k}\cdot
  \mathbf{r}_j}+ \textrm{h.c.}\label{s3bis}\\
  \frac{d\hat a_{\mathbf{k}}}{dt} &=& \frac{1}{i\hbar}[\hat a_{\mathbf{k}},\hat H]=
  -ig_ke^{i(\omega_k-\omega_0)t}\sum_{m=1}^N \hat\sigma_{m}e^{-i\mathbf{k}\cdot
  \mathbf{r}_m}\label{akbis}.
\end{eqnarray}
We consider the atoms initially in their ground state and we
assume weak excitation ($\Omega_0\ll\Gamma$), so that we
approximate $\hat\sigma_{3j}(t)\approx -\hat I_j$, where $\hat I_j$ is
the identity operator for the $j$th atom. This approximation amounts to neglecting
saturation and multiexcitation, i.e. all the processes generating
more than one photon at the same time (\textit{linear regime}).
Integrating Eq.(\ref{akbis}) and substituting it into
Eq.(\ref{s1bis}), neglecting $a_k(0)$ (since the initial field
state is vacuum) we obtain
\begin{eqnarray}
  \frac{d\hat\sigma_{j}}{dt} &=&
  \left(i\Delta_0-\frac{\Gamma}{2}\right)\hat\sigma_j-\frac{i\Omega_0}{2} \hat I_j e^{i\mathbf{k}_0\cdot\mathbf{r}_j}-
  \sum_{\mathbf{k}}g_k^2\sum_{m=1}^N
  e^{i\mathbf{k}\cdot(\mathbf{r}_j-\mathbf{r}_m)}
  \int_0^t dt'
  \hat\sigma_m(t-t')\,e^{-i(\omega_k-\omega_0)t'}\label{s1ter}.
\end{eqnarray}
The last term in Eq.(\ref{s1ter}) describes the effect of the
spontaneously emitted photons on the atoms. In the Markov approximation (i.e.
when the photon transit time through the atomic sample is much shorter than the excitation decay time), we may approximate under
the integral $\hat\sigma_m(t-t')\approx \hat\sigma_m(t)$.
Then, the remaining time integral yields a real part [with a term  proportional to $\delta (k-k_0)$]
and an imaginary part (corresponding to the principal part of the integral).
We transform the sum over the modes $\mathbf{k}$ into an integral,
$\sum_{\mathbf{k}}\rightarrow (V_{ph}/8\pi^3)\int d\mathbf{k}$.
The real and imaginary parts of the double integral over $t$
and $\mathbf{k}$ yield the cooperative decay and frequency shift
(collective Lamb shift), respectively. The proper expression of
the cooperative frequency shift has been obtained  adding to the
Hamiltonian (\ref{H1}) the not-RWA contributions associated to
virtual photons exchanged between different atoms. It results in the
following relation \cite{Svidzinsky2010}:
\begin{equation}\label{rel:kern}
\sum_{\mathbf{k}}g_k^2
  e^{i\mathbf{k}\cdot
   \mathbf{R}}
  \int_0^\infty dt'e^{-ic(k-k_0)t'}\longrightarrow \frac{\Gamma}{2ik_0
  |\mathbf{R}|}e^{ik_0 |\mathbf{R}|}
\end{equation}
where $\Gamma=V_{ph}g_{k_0}^2 k_0^2/(\pi c)$.
Using Eq.(\ref{rel:kern}) in Eq.(\ref{s1ter}) we obtain
\cite{JMO2011},
\begin{eqnarray}
   \frac{d\hat\sigma_{j}(t)}{dt} &=&
  \left(i\Delta_0-\frac{\Gamma}{2}\right)\hat\sigma_j(t)-\frac{i\Omega_0}{2}\hat I_j e^{i\mathbf{k}_0\cdot\mathbf{r}_j}-\frac{\Gamma}{2}
  \sum_{m\neq j}^N\gamma_{jm}
  \hat\sigma_m(t)\label{s1:Mark}.
\end{eqnarray}
where $\gamma_{jm}=\exp(ik_0r_{jm})/(ik_0 r_{jm})$ and $r_{jm}=|\mathbf{r}_j-\mathbf{r}_m|$. Equation (\ref{s1:Mark})
describes the time evolution of the atomic operators $\hat\sigma_{j}$ of $N$ weakly excited atoms. The real part of $\gamma_{jm}$ describes the
spontaneous emission decay and the imaginary part of $\gamma_{jm}$
describes the energy shift due to resonant dipole-dipole
interactions.
Note that we use a scalar model for the field, neglecting thus any polarization and near field dependence.
Detailed calculations for small and large samples of various geometries show that near-field and far-field contributions as well
as resonant and antiresonant terms need to be taken properly into account for quantitative predictions \cite{Friedberg1973,Friedberg2010},
and the present model thus needs to be considered with care illustrating only a part of the dipole-dipole coupling for real systems.

The positive-frequency part of the electric field is defined as
\begin{equation}\label{E}
    \hat E^{(+)}(\mathbf{r},t)=i\sum_{\mathbf{k}}{\cal
    E}_{k}\hat a_\mathbf{k}(t) e^{-i\omega_k t+i\mathbf{k}\cdot\mathbf{r}}
\end{equation}
where
${\cal E}_{k}=\sqrt{\hbar\omega_k/2\epsilon_0 V_{ph}}$
is the
single-photon electric field. By integrating Eq.(\ref{akbis}) and
inserting it in Eq.(\ref{E}) we obtain
\begin{equation}\label{E2}
    \hat E^{(+)}(\mathbf{r},t)=\sum_{\mathbf{k}}{\cal
    E}_{k}g_k\sum_{m=1}^N e^{i\mathbf{k}\cdot(\mathbf{r}-\mathbf{r}_m)-i\omega_0 t}
    \int_0^t dt'e^{-i(\omega_k-\omega_0)t'}\hat\sigma_m(t-t')
\end{equation}
Using Eq.(\ref{rel:kern}), the Markov approximation leads to
\begin{equation}\label{E3}
    \hat E^{(+)}(\mathbf{r},t)\approx -i\frac{d k_0^2}{4\pi\epsilon_0}\sum_{j=1}^N
    \frac{e^{-i\omega_0(t-|\mathbf{r}-\mathbf{r}_j|/c)}}{|\mathbf{r}-\mathbf{r}_j|}
    \hat\sigma_j(t)
\end{equation}
which has a transparent interpretation as the sum of wavelets
scattered by $N$ dipoles of position $\mathbf{r}_j$ and detected
at distance $\mathbf{r}$ and time $t$. In the far field
limit, $|\mathbf{r}-\mathbf{r}_j|\approx r-(\mathbf{r}\cdot
\mathbf{r}_j)/r$ and \cite{Rehler1971}
\begin{equation}\label{E4}
    \hat E^{(+)}(\mathbf{r},t)\approx -i\frac{d k_0^2}{4\pi\epsilon_0 r}e^{-i\omega_0(t-r/c)}\sum_{j=1}^N
    e^{-i\mathbf{k}\cdot \mathbf{r}_j}
    \hat\sigma_j(t)
\end{equation}
where $\mathbf{k}=k_0(\mathbf{r}/r)$. The intensity of scattered radiation by $N$ atoms measured at distance
$\mathbf{r}$ and time $t$ is
\begin{equation}\label{G1:def}
    I_N(\mathbf{r},t,\mathbf{r},t)=\frac{c\epsilon_0}{2}\langle\hat E^{(-)}(\mathbf{r},t)\hat E^{(+)}(\mathbf{r},t)\rangle=\frac{cd^2k_0^4}{32\pi^2\epsilon_0 r^2}\sin^2\theta_{\mathbf{k}}\cdot I(t)   
\end{equation}
where $\theta_{\mathbf{k}}$ is the detection angle and
\begin{equation}
I(t)=\overline{\sum_{j,m}e^{-i\mathbf{k}\cdot(\mathbf{r}_j-\mathbf{r}_m)}\langle\hat\sigma_m^\dagger(t)\hat\sigma_j(t)\rangle}    
\end{equation}
is the dimensionless intensity, where the bar calls for an average over positions, assuming that the locations
of the $N$ atoms are not controlled. In that case an average over a series of experiments with
many otherwise identical samples will be needed to lead to a stable observed intensity.

\section{Correlation functions}

The first-order and second-order coherence of light can be described by the normalized two-time and equal position correlation functions introduced by Glauber \cite{Glauber1963a,Glauber1963b}:
\begin{eqnarray}\label{gg}
  g^{(1)}(t,\tau) &=& \frac{\langle E^{(-)}(\mathbf{r},t)E^{(+)}(\mathbf{r},t+\tau)\rangle}
  {\langle E^{(-)}(\mathbf{r},t)E^{(+)}(\mathbf{r},t)\rangle}\\
  g^{(2)}(t,\tau) &=& \frac{\langle E^{(-)}(\mathbf{r},t)E^{(-)}(\mathbf{r},t+\tau)
  E^{(+)}(\mathbf{r},t+\tau)E^{(+)}(\mathbf{r},t)\rangle}
  {\langle E^{(-)}(\mathbf{r},t)E^{(+)}(\mathbf{r},t)\rangle^2}
\end{eqnarray}
(we drop the hats over the operators) where $\tau$ is the time difference between two-photon detection events within a two-photon
coincidence count. Two-photon bunching is defined as $g^{(2)}(0)>g^{(2)}(\tau)$ ($\tau\neq 0$), whereas antibunching is defined as $g^{(2)}(0)<g^{(2)}(\tau)$ ($\tau\neq 0$), which is usually regarded as a nonclassical effect \cite{Mandel}.
Using (\ref{E4}) we obtain
\begin{eqnarray}\label{gg:2}
  g^{(1)}(t,\tau) &=& \frac{1}{I(t)}\overline{\sum_{j,m}
  e^{-i\mathbf{k}\cdot(\mathbf{r}_j-\mathbf{r}_m)}\langle\sigma_m^\dagger(t)\sigma_j(t+\tau)\rangle}\\
  g^{(2)}(t,\tau) &=& \frac{1}{I^2(t)}\overline{\sum_{j,m,p,q}
  e^{-i\mathbf{k}\cdot(\mathbf{r}_j-\mathbf{r}_m+\mathbf{r}_q-\mathbf{r}_p)}
    \langle\sigma_p^\dagger(t)\sigma_m^\dagger(t+\tau)\sigma_j(t+\tau)\sigma_q(t)\rangle}.
\end{eqnarray}
The determination of the
correlation functions $g^{(1)}(t,\tau)$ and $g^{(2)}(t,\tau)$
requires the evaluation of the two-time quantum averages
$\langle\sigma_m^\dagger(t)\sigma_j(t+\tau)\rangle$ and
$\langle\sigma_p^\dagger(t)\sigma_m^\dagger(t+\tau)\sigma_j(t+\tau)\sigma_q(t)\rangle$,
which can be related to that of averages evaluated at a single
time using the quantum regression theorem \cite{Loudon,Barnett}.

\section{Single-excitation state}\label{Single}

Let us restrict the Hilbert space of the $N$ atoms to the subspace spanned by the ground state
$|g\rangle=|g_1,\dots,g_N\rangle$ and the single-excited-atom
states $|j\rangle=|g_1,\dots,e_j,\dots,g_N\rangle$ with
$j=1,\dots,N$:
\begin{equation}\label{psi}
    |\Psi(t)\rangle=\alpha(t)|g\rangle+e^{-i\Delta_0
    t}\sum_{j=1}^N\beta_j(t)|j\rangle
\end{equation}
where we approximate $\alpha\approx 1$, so that $\langle\hat\sigma_j\rangle=\alpha^*\beta_j\approx \beta_j$ where, from Eq.(\ref{s1:Mark}),
\begin{eqnarray}
   \frac{d\beta_{j}(t)}{dt} &=&
  \left(i\Delta_0-\frac{\Gamma}{2}\right)\beta_j(t)-\frac{i\Omega_0}{2}e^{i\mathbf{k}_0\cdot\mathbf{r}_j}-\frac{\Gamma}{2}
  \sum_{m\neq j}\gamma_{jm}
  \beta_m(t)\label{s1:beta}
\end{eqnarray}
with initial conditions $\beta_j(0)=0$. Since $\langle\sigma_m^\dagger(t)\sigma_j(t)\rangle =\beta_m^*(t)\beta_j(t)$, the dimensionless average intensity at the time $t$ is
\begin{eqnarray}
  I(t)&=&
    \overline{\left|\sum_j e^{-i
    \mathbf{k}\cdot\mathbf{r}_j}\beta_j(t)\right|^2}.\label{I(t)}
\end{eqnarray}
Using the quantum regression theorem, it is possible to show that (see Appendix \ref{g1:g2})
\begin{eqnarray}\label{gg:1:gen}
  g^{(1)}(t,\tau) &=& \frac{1}{I(t)}\overline{\sum_{j,m}e^{-i\mathbf{k}\cdot(\mathbf{r}_j-\mathbf{r}_m)}\beta_m^*(t)\beta_j(t+\tau)}e^{-i\omega_0\tau}.
\end{eqnarray}
and
\begin{eqnarray}\label{gg:2:fin}
  g^{(2)}(t,\tau) &=& \frac{1}{I^2(t)}\overline{\sum_{j,m,p,q}e^{-i\mathbf{k}\cdot(\mathbf{r}_j-\mathbf{r}_m+\mathbf{r}_q-\mathbf{r}_p)}
    \beta_p^*(t)H_{mj}(\tau)\beta_q(t)}
\end{eqnarray}
where $H_{mj}(\tau)$ is the solution of the equation
\begin{eqnarray}
  \frac{d}{d\tau}H_{mj}(\tau)&=&-\Gamma H_{mj}(\tau)-
  \frac{i\Omega_0}{2}\left[e^{i\mathbf{k}_0\cdot\mathbf{r}_j}\beta^*_m(\tau)
  -e^{-i\mathbf{k}_0\cdot\mathbf{r}_m}\beta_j(\tau)\right]
  -\frac{\Gamma}{2}
  \left[\sum_{k\neq j}\gamma_{jk}H_{mk}(\tau)+\sum_{k\neq m}\gamma^*_{mk}H_{kj}(\tau)\right]\label{Hmj}.
\end{eqnarray}
with $H_{mj}(0)=0$. We observe that, since $g^{(1)}(t,0)=1$ and $\lim_{t\rightarrow\infty}|g^{(1)}(t,\tau)|= 1$,
the system has full first-order coherence. We can interpret this result also saying that the scattering is elastic, as expected by a system with a linear response to a cw excitation. In this case the randomness of the system has no effect on the field correlation function.
Instead, $g^{(2)}(t,0)=0$ since the state (\ref{psi}) has a single excitation and it is not possible to detect two photons at the same time. The numerical solution of Eq.(\ref{gg:2:fin}) will be discussed in Sec. \ref{numerics}.

\section{Timed Dicke state}\label{Dicke}

\begin{figure}[h]
        \centerline{\scalebox{0.35}{\includegraphics{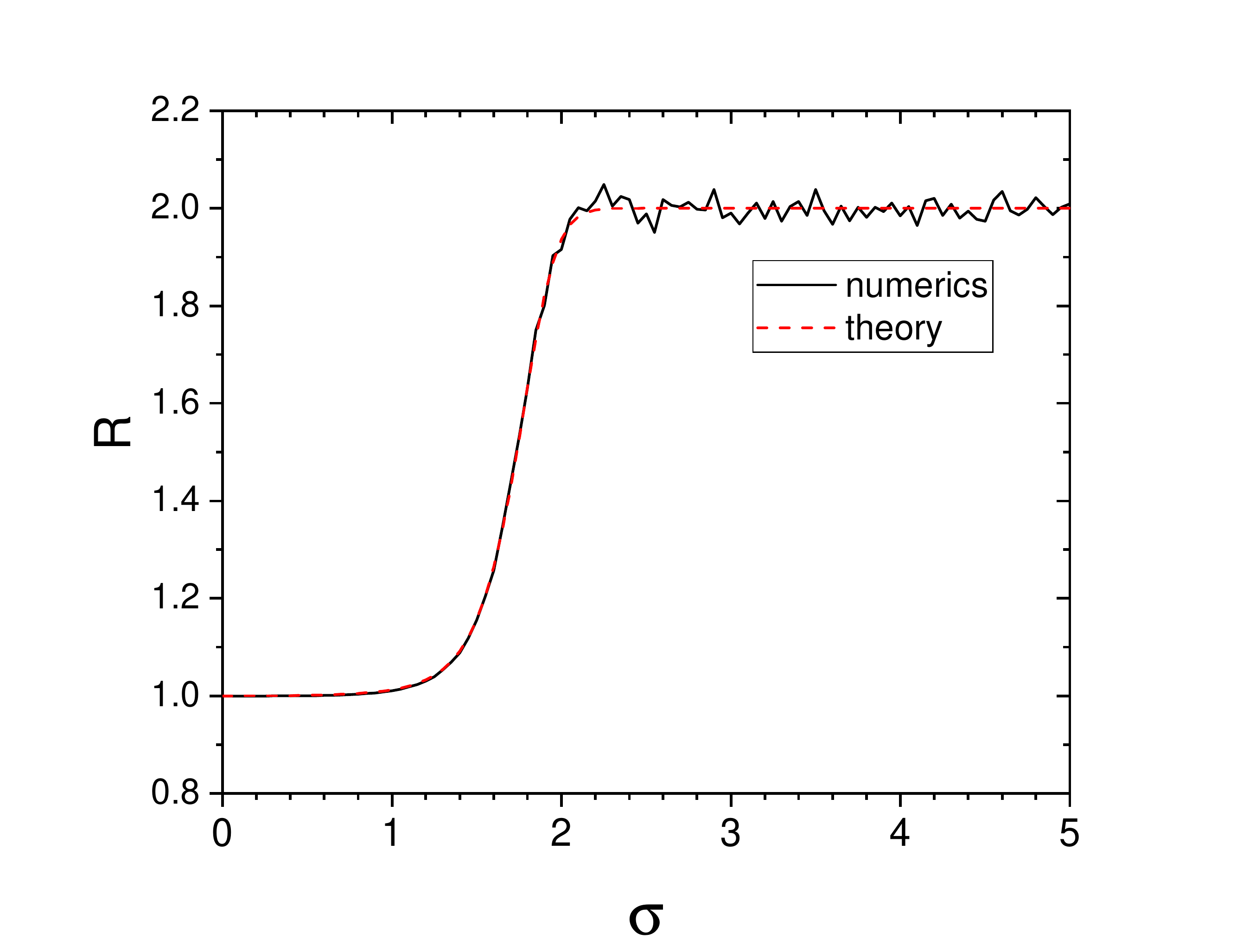}}}
        \caption{$R$ vs $\sigma=k_0\sigma_r$ for $N=10^3$ and $\theta=90^\circ$, averaged over $10^4$ iterations; solid black line: numerical simulation; 
        dashed red line: analytic solution.}
        \label{fig1}
\end{figure}

Let's now consider the timed Dicke state of Eq. (\ref{TDS}), assuming
$\beta_j(t)=\beta(t)e^{i\mathbf{k}_0\cdot \mathbf{r}_j}$ in Eq. (\ref{psi}). Then,
\begin{equation}
I(t)=N^2\overline{|S_N(\mathbf{k}-\mathbf{k}_0)|^2}|\beta(t)|^2
\end{equation}
 where
\begin{equation}\label{SN}
    S_N(\mathbf{k}-\mathbf{k}_0)=\frac{1}{N}\sum_{j=1}^N
e^{-i(\mathbf{k}-\mathbf{k}_0)\cdot
\mathbf{r}_j}
\end{equation}
is the structure factor. Furthermore, since the
dynamics is the same for all the atoms, with $\langle\sigma_j(t)\rangle=\langle\sigma(t)\rangle e^{i\mathbf{k}_0\cdot \mathbf{r}_j}$,
\begin{eqnarray}\label{gg:TDS}
  g^{(1)}(t,\tau) &=& \frac{\langle\sigma^\dagger(t)\sigma(t+\tau)\rangle}
  {\langle\sigma^\dagger(t)\sigma(t)\rangle}e^{-i\omega_0\tau}\\
  g^{(2)}(t,\tau) &=&R
  \frac{
    \langle\sigma^\dagger(t)\sigma^\dagger(t+\tau)\sigma(t+\tau)\sigma(t)\rangle}
  {\langle\sigma^\dagger(t)\sigma(t)\rangle^2}
\end{eqnarray}
where the factor $R$ is (assuming $N\gg 1$)
\begin{equation}
R=\frac{\overline{|S_N|^4}}{\left[\overline{|S_N|^2}\right]^2}\approx 
\frac{2+4N|S_\infty(\mathbf{k}-\mathbf{k}_0)|^2+N^2|S_\infty(\mathbf{k}-\mathbf{k}_0)|^4}
{(1+N|S_\infty(\mathbf{k}-\mathbf{k}_0)|^2)^2}
\end{equation}
and we have approximated the structure factor by a continuous distribution with number density $n(\mathbf{r})$,
\begin{equation}
S_\infty(\mathbf{k}-\mathbf{k}_0)=\int  n(\mathbf{r}) e^{-i(\mathbf{k}-\mathbf{k}_0)\cdot
\mathbf{r}_j}d\mathbf{r}.
\end{equation}
The term $R$ describes the classical intensity correlation function at zero delay, due to the spatial distribution of the scatterers. It ranges from $R=1$ to $R=2$ for coherent and chaotic light, respectively.
For instance, for a Gaussian spherical distribution with rms size $\sigma_r$, $S_\infty(\mathbf{k}-\mathbf{k}_0)=\exp[-2(k_0\sigma_r)^2\sin^2(\theta/2)]$. Hence $R=1$ for $k_0\sigma_r|\sin(\theta/2)|\ll 1$ and $R=2$ for $k_0\sigma_r|\sin(\theta/2)|\gg 1$ (see Fig. \ref{fig1}).

\begin{figure}[h]
        \centerline{\scalebox{0.35}{\includegraphics{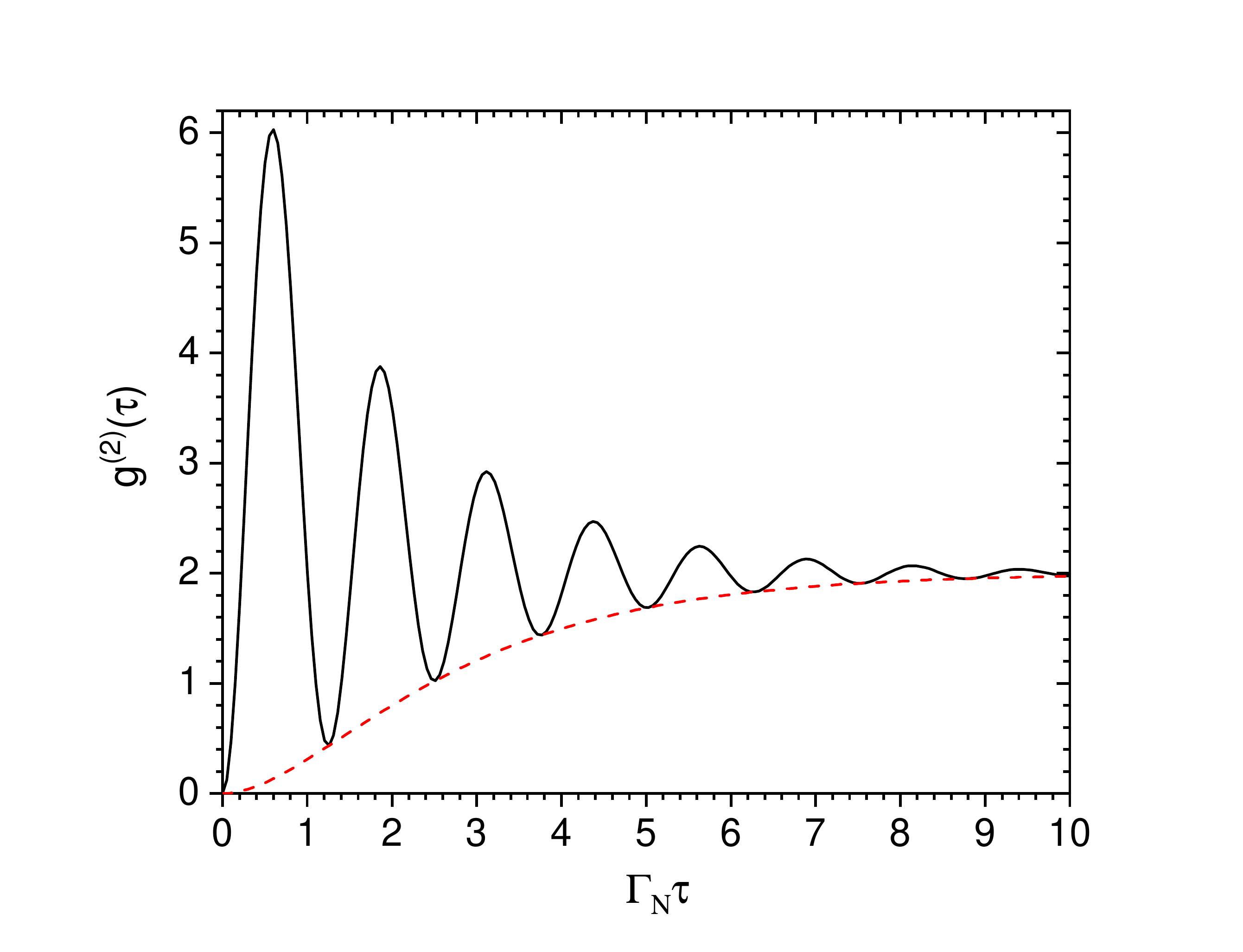}}}
        \caption{$g^{(2)}(\tau)$ vs $\Gamma_N	\tau$ for $R=2$ and two different values of detuning: $\Delta=5\Gamma_N$ (solid black line) and $\Delta=0$ (red dashed line). }
        \label{fig2}
    \end{figure}

For a timed Dicke state, $\langle\sigma^\dagger(t)\sigma(t)\rangle=|\beta(t)|^2$, with
\begin{equation}
\beta(t)=\beta^{st}\left\{1-e^{(i\Delta-\Gamma_N/2)t}\right\},\label{beta:TD}
\end{equation}
\begin{equation}
\beta^{st}=\frac{\Omega_0}{2\Delta+i\Gamma_N},
\end{equation}
$\Gamma_N=\Gamma+(N\Gamma/4\pi)\int_0^{2\pi}d\phi\int_0^\pi 
d\theta\sin\theta|S_\infty(k_0,\theta,\phi)|^2$ is the collective decay rate,  $\Delta=\Delta_0-\Delta_N$ and 
$\Delta_N=(N\Gamma/8\pi^2)\mathsf{P}\int_0^\infty d\kappa\kappa^3/(\kappa-1)\int_0^{2\pi}d\phi\int_0^\pi 
d\theta\sin\theta|S_\infty(k_0\kappa,\theta,\phi)|^2$ is the collective Lamb shift \cite{JMO2011}. 
Since
$\langle\sigma(t)\rangle=\beta(t)$, then
\begin{equation}\label{beta:tau}
    \langle\sigma(t+\tau)\rangle=\beta(t+\tau)=\beta^{st}+[\beta(t)-\beta^{st}]e^{(i\Delta-\Gamma_N/2)\tau}
    =\langle\sigma(\tau)\rangle+e^{(i\Delta-\Gamma_N/2)\tau}\langle\sigma(t)\rangle.
\end{equation}
From the quantum regression theorem,
\begin{equation}\label{QRT:1}
\langle\sigma^\dagger(t)\sigma(t+\tau)\rangle=
\langle\sigma^\dagger(t)\rangle\langle\sigma(\tau)\rangle+e^{(i\Delta-\Gamma_N/2)\tau}\langle\sigma^\dagger(t)\sigma(t)\rangle
\end{equation}
and
\begin{equation}\label{g1:beta}
 g^{(1)}(t,\tau)=\frac{\beta^*(t)\beta(\tau)+e^{(i\Delta-\Gamma_N/2)\tau}|\beta(t)|^2}{|\beta(t)|^2}e^{-i\omega_0\tau}=
 \frac{1-e^{(i\Delta-\Gamma_N/2)(t+\tau)}}{1-e^{(i\Delta-\Gamma_N/2)t}}e^{-i\omega_0\tau}.
\end{equation}
As expected, $\lim_{t\rightarrow\infty}|g^{(1)}(t,\tau)|= 1$ and the scattered field has full first-order coherence.
In order to evaluate $g^{(2)}(t,\tau)$, we note that
\begin{eqnarray}\label{beta:tau}
    \langle\sigma^\dagger(t+\tau)\sigma(t+\tau)\rangle&=&|\beta(t+\tau)|^2\nonumber\\
    &=&\left|\beta^{st}+[\beta(t)-\beta^{st}]e^{(i\Delta-\Gamma_N/2)\tau}\right|^2\nonumber\\
    &=&|\beta^{st}|^2+|\beta(t)-\beta^{st}|^2e^{-\Gamma_N\tau}
    + 
    \left\{
    (\beta(t)-\beta^{st}){\beta^{st}}^*e^{i\Delta\tau}+\mathrm{c.c.}
    \right\}e^{-\Gamma_N\tau/2}.
\end{eqnarray}
where $\mathrm{c.c.}$ stands for the complex conjugate.
It can be rewritten in the form
\begin{eqnarray}\label{beta:tau:2}
    \langle\sigma^\dagger(t+\tau)\sigma(t+\tau)\rangle&=&A(\tau)+B(\tau)\langle\sigma(t)\rangle+B^*(\tau)\langle\sigma^\dagger(t)\rangle
    + C(\tau)\langle\sigma^\dagger(t)\sigma(t)\rangle
\end{eqnarray}
where
\begin{eqnarray}\label{coeff}
    A(\tau)&=&|\beta^{st}|^2\left\{1+e^{-\Gamma_N\tau}-2e^{-\Gamma_N\tau/2}\cos(\Delta\tau)\right\}\label{A}\\
    B(\tau)&=&e^{-\Gamma_N\tau/2}{\beta^{st}}^*\left\{e^{i\Delta\tau}-e^{-\Gamma_N\tau/2}\right\}\label{B}\\
    C(\tau)&=&e^{-\Gamma_N\tau}\label{C}.
\end{eqnarray}
From the quantum regression theorem,
\begin{eqnarray}\label{QRT:2}
\langle\sigma^\dagger(t)\sigma^\dagger(t+\tau)\sigma(t+\tau)\sigma(t)\rangle&=&
A(\tau)\langle\sigma^\dagger(t)\sigma(t)\rangle
+B(\tau)\langle\sigma^\dagger(t)\sigma(t)\sigma(t)\rangle
+B^*(\tau)\langle\sigma^\dagger(t)\sigma^\dagger(t)\sigma(t)\rangle\nonumber\\
&+&C(\tau)\langle\sigma^\dagger(t)\sigma^\dagger(t)\sigma(t)\sigma(t)\rangle.
\end{eqnarray}
Since $\sigma(t)\sigma(t)=0$ and
$\sigma^\dagger(t)\sigma^\dagger(t)=0$,
\begin{equation}\label{g2:beta}
 g^{(2)}(t,\tau)=R\frac{A(\tau)}{|\beta(t)|^2}=
 R\frac{1+e^{-\Gamma_N\tau}-2e^{-\Gamma_N\tau/2}\cos(\Delta\tau)}
 {1+e^{-\Gamma_N t}-2e^{-\Gamma_N t/2}\cos(\Delta t)}
\end{equation}
Hence, $\lim_{t\rightarrow\infty} g^{(2)}(t,\tau)= g^{(2)}(\tau)$ where
\begin{equation}\label{g2:fin}
 g^{(2)}(\tau)=R\left|1-e^{(i\Delta-\Gamma_N/2)\tau}\right|^2=
 R\left\{1+e^{-\Gamma_N\tau}-2e^{-\Gamma_N\tau/2}\cos(\Delta\tau)\right\}.
\end{equation}
In conclusion, the timed Dicke state behaves as a single driven
atom, but with a collective decay rate $\Gamma_N$ and Lamb shift $\Delta_N$. As
expected, $g^{(2)}(0)=0$ and $\lim_{\tau\rightarrow\infty}g^{(2)}(\tau)= R$. Figure \ref{fig2} shows Eq.(\ref{g2:fin}) for $R=2$ and two different values of detuning, $\Delta=5\Gamma_N$ (solid black line) and $\Delta=0$ (red dashed line).
Since $g^{(2)}(\tau)>g^{(2)}(0)$, the system exhibits
antibunching. Notice that, contrary to $g^{(1)}$,  $g^{(2)}$ is proportional to the factor $R$, equal to 2 when the photons are emitted isotropically and randomly out of the diffraction cone, with aperture $\Delta\theta\sim \lambda_0/\sigma_r$, and equal to 1 in the opposite case.
In the case of the timed Dicke state, $H_{mj}(\tau)=H(\tau)$ and  $g^{(2)}(\tau)=H(\tau)/|\beta^{st}|^2$, where
$H(\tau)$ is the solution of the equation
\begin{eqnarray}
  \frac{dH(\tau)}{d\tau}&=&
  \frac{i\Omega_0}{2}[\beta(\tau)-\beta^*(\tau)]
  -\Gamma_N H(\tau)\label{H},
\end{eqnarray}
yielding the solution (\ref{g2:fin}).

\section{Product state}

It has been noted that the results for the intensity $I(t)$ emitted by $N$ weakly-excited  two-level atoms, obtained for a symmetric timed Dicke state as described in Sec.\ref{Dicke}, could be obtained assuming a product state of $N$ two-level atoms \cite{Eberly2006} (named also ‘pure Bloch state’ by some authors \cite{Friedberg2007}). More specifically, pure Bloch states are product states in which every one of $N$ two-level atoms is in the same superposition of ground and excited state. Such states are easily produced experimentally. As it happens for the timed Dicke state, the driving field imposes a coherence in the photons emitted spontaneously by each atom, so that superradiance arises because the state is symmetric under exchange of particles. However, it is expected that the quantum statistic of the timed Dicke state will be quite different from that of the pure Bloch state. The aim of this section is to obtain the stationary correlation functions $g^{(1)}(\tau)$ and $g^{(2)}(\tau)$ for such product state, in the presence of a cw driving field. To be more general, we assume first that the excitation probability amplitudes $\beta_j$ are not the same for every atom, 
\begin{equation}\label{state:product}
|\Psi(t)\rangle=\prod_{j=1}^N \left\{
\alpha(t)|g_j\rangle+\beta_j(t)e^{-i\Delta_0 t}|e_j\rangle \right\}
\end{equation}
with $|\alpha(t)|^2+|\beta_j(t)|^2=1$ for every $j$. Assuming $\alpha\sim 1$, we obtain $\langle\sigma_j\rangle=\beta_j$, $\langle\sigma_m^\dagger\sigma_j\rangle=\beta_m^*\beta_j$ and 
\begin{equation}
\langle\sigma_m\sigma_j\rangle
= \left\{
\begin{array}{ccc} \displaystyle
 \beta_m\beta_j & \mathrm{if} & j\neq m \\
 & & \\
 0 & \mathrm{if} & j=m
\end{array}\right..
\end{equation}
Hence, for the product state all the expectation values of the operators factorize and the dynamics is determined solely by $\beta_j$, solution of Eq.(\ref{s1:beta}). As a consequence, following the same reasoning adopted in Sec.\ref{Single}, we arrive at the same expression (\ref{gg:1:gen}) for $g^{(1)}(t,\tau)$ obtained assuming the entangled state (\ref{psi}). Differences between the product state and the single-excitation state appear when higher-order expectation values are observed. We report the details of the demonstration in Appendix \ref{product}, leading to
\begin{eqnarray}\label{gg:2:prod}
  g^{(2)}(t,\tau) &=& \frac{1}{I^2(t)}\overline{\sum_{j,m,p,q}e^{-i\mathbf{k}\cdot(\mathbf{r}_j-\mathbf{r}_m+\mathbf{r}_q-\mathbf{r}_p)}
    \beta_p^*(t)H_{mj}(t,\tau)\beta_q(t)}
\end{eqnarray}
where $H_{mj}(\tau)$ is the solution of the equation
\begin{eqnarray}
  \frac{d}{d\tau}H_{mj}(t,\tau)&=&-\Gamma H_{mj}(t,\tau)-
  \frac{i\Omega_0}{2}\left[e^{i\mathbf{k}_0\cdot\mathbf{r}_j}F^*_m(t,\tau)
  -e^{-i\mathbf{k}_0\cdot\mathbf{r}_m}F_j(t,\tau)\right]
  -\frac{\Gamma}{2}
  \left[\sum_{k\neq j}\gamma_{jk}H_{mk}(t,\tau)+\sum_{k\neq m}\gamma^*_{mk}H_{kj}(t,\tau)\right]\nonumber\\
  \label{Hmj:prod}
\end{eqnarray}
with initial condition
\begin{equation}
H_{mj}(t,0)
= \left\{
\begin{array}{ccc} \displaystyle
 \beta_m^*(t)\beta_j(t) & & \mathrm{if}\quad p\neq m\quad \mathrm{or}\quad  j\neq q\\
 & & \\
 0 &   & \mathrm{otherwise}
\end{array}\right.
\end{equation}
and $F_j(t,\tau)$ is the solution of
\begin{eqnarray}
  \frac{d}{d\tau}F_{j}(t,\tau)&=&\left(i\Delta_0-\frac{\Gamma}{2}\right)F_{j}(t,\tau)-
  \frac{i\Omega_0}{2}e^{i\mathbf{k}_0\cdot\mathbf{r}_j}-\frac{\Gamma}{2}
  \sum_{k\neq j}\gamma_{jk}F_{k}(t,\tau)\label{j}.
\end{eqnarray}
with 
\begin{equation}
F_{j}(t,0)
= \left\{
\begin{array}{ccc} \displaystyle
\beta_j(t) & \mathrm{if} & j\neq q \\
 & & \\
 0 &  & \mathrm{otherwise}
\end{array}\right..
\end{equation}
The numerical solution of Eq. (\ref{gg:2:prod}) will be discussed in Sec. \ref{numerics}.

\section{The Eberly's state}

\begin{figure}[h]
        \centerline{\scalebox{0.5}{\includegraphics{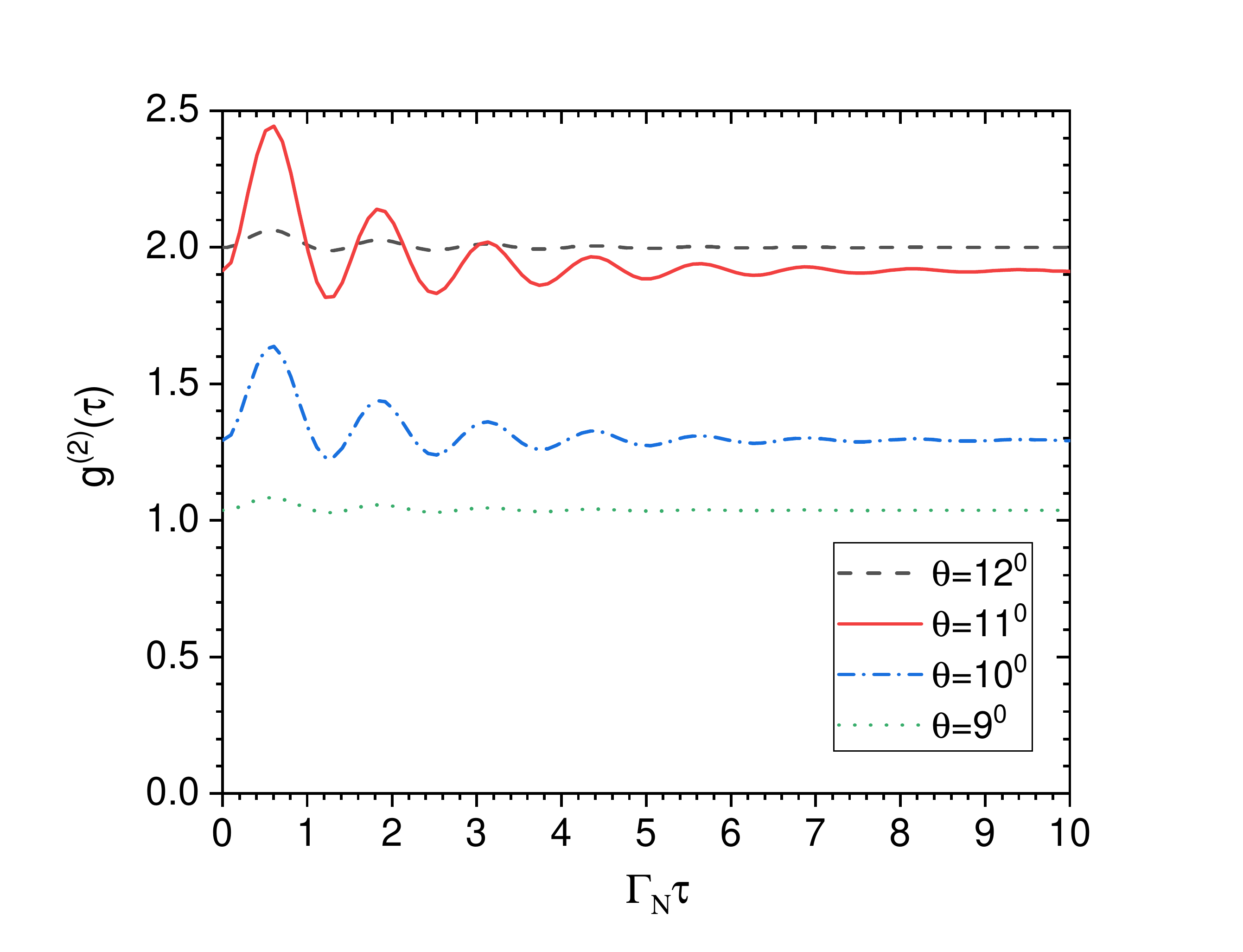}}}
        \caption{$g^{(2)}(\tau)$ vs $\Gamma_N\tau $ from Eq.(\ref{g2:p}) for $\Delta=5\Gamma_N$, $N=10^6$, $k_0\sigma_r=20$ and angles $\theta=12^\circ$ (dashed black line), $\theta=11^\circ$ (continuous red line), $\theta=10^\circ$ (dashed-dotted blue line) and $\theta=9^\circ$ (dotted green line).}
        \label{fig3}
    \end{figure}

As done for the timed Dicke state, we assume now that the excitation probability is the same for all the atoms \cite{Eberly2006},
\begin{equation}\label{state:Eberly}
|\Psi(t)\rangle_{E}=\prod_{j=1}^N \left\{
\alpha(t)|g_j\rangle+\beta(t)e^{-i\Delta_0 t+i \mathbf{k}_0\cdot
\mathbf{r}_j}|e_j\rangle \right\}
\end{equation}
with $|\alpha(t)|^2+|\beta(t)|^2=1$. In the following we will
assume the linear approximation, so that $\alpha(t)\sim 1$ and
$\beta(t)$ is given by Eq.(\ref{beta:TD}). Then, defining
\begin{equation}\label{EE}
    E(t)=\frac{1}{N}\sum_{j=1}^N e^{-i\mathbf{k}\cdot\mathbf{r}_j}\sigma_{j}(t)
\end{equation}
we have
\begin{eqnarray}
  \langle E(t)\rangle &=& S_N\beta(t) \\
  \langle E(t+\tau)\rangle &=&  S_N\beta(t+\tau)=S_N\beta(\tau)+e^{(i\Delta-\Gamma_N/2)\tau}\langle E(t)\rangle\\
  \langle E^\dagger(t)E(t)\rangle &=& |S_N|^2|\beta(t)|^2\\
  \langle E^\dagger(t+\tau)E(t+\tau)\rangle &=&
  A(\tau)|S_N|^2 +B(\tau)S_N^*\langle E(t)\rangle+B^*(\tau)S_N\langle E^\dagger(t)\rangle
    + C(\tau)\langle E^\dagger(t) E(t)\rangle\label{E+E}
\end{eqnarray}
where $S_N$ has been defined in Eq.(\ref{SN}) and $A(\tau)$, $B(\tau)$ and $C(\tau)$ have been defined in
Eqs.(\ref{A})-(\ref{C}). From the quantum linear regression
theorem,
\begin{eqnarray}
  \langle E^\dagger(t)E(t+\tau)\rangle &=& S_N\beta(\tau)\langle E^\dagger(t)\rangle
  +e^{(i\Delta-\Gamma_N/2)\tau}\langle E^\dagger(t) E(t)\rangle\nonumber\\
\end{eqnarray}
and
\begin{eqnarray}
 g^{(1)}(t,\tau) &=&
  \frac{\beta(\tau)\beta^*(t)+e^{(i\Delta-\Gamma_N/2)\tau}|\beta(t)|^2}{|\beta(t)|^2}e^{-i\omega_0\tau}
\end{eqnarray}
which coincides with Eq.(\ref{g1:beta}). From Eq.(\ref{E+E}) we
obtain
\begin{eqnarray}
  \langle E^\dagger(t)E^\dagger(t+\tau)E(t+\tau)E(t)\rangle &=&
  A(\tau)|S_N|^2\langle E^\dagger(t) E(t)\rangle\nonumber\\
  &+&\left[B(\tau)S_N^*\langle E^\dagger(t)E(t) E(t)\rangle+\textrm{c.c.}\right]\nonumber\\
  &+& C(\tau)\langle E^\dagger(t)E^\dagger(t)E(t) E(t)\rangle\label{EEEE}
\end{eqnarray}
The last two terms make the difference from the timed Dicke state,
since in general they are not zero. It is possible to demonstrate that (see Appendix \ref{Eberly})
\begin{eqnarray}
  \langle E^\dagger(t)E(t) E(t)\rangle &=&
    \beta(t)|\beta(t)|^2\left\{
    \frac{2(N-1)}{N^2}+K_2\right\}\label{EEE:1}\\
  \langle E^\dagger(t)E^\dagger(t)E(t) E(t)\rangle &=& |\beta(t)|^4\left\{
    \frac{2(N-1)}{N^3}+\frac{4(N-2)}{N^2}K_2+K_4\right\}\label{EEEE:1}
\end{eqnarray}
where
\begin{eqnarray}
  K_2 &=& \frac{1}{N^2}\sum_{j}\sum_{p\neq j} e^{i(\mathbf{k}_0-
    \mathbf{k})\cdot(\mathbf{r}_p-\mathbf{r}_j)}\approx|S_\infty|^2\\
     K_4 &=&  \frac{1}{N^4}\sum_{m}\sum_{j\neq m}\sum_{p\neq j,m} \sum_{q\neq p,j,m}
    e^{i(\mathbf{k}_0-
    \mathbf{k})\cdot(\mathbf{r}_p+\mathbf{r}_q-\mathbf{r}_j-\mathbf{r}_m)}\approx|S_\infty|^4
\end{eqnarray}
For $N\gg 1$ and after the statistical average,
\begin{eqnarray}
  \langle E^\dagger(t)E^\dagger(t+\tau)E(t+\tau)E(t)\rangle &=&
  \left(
    \frac{2}{N^2}+\frac{4}{N}|S_\infty|^2+|S_\infty|^4\right)\left(A(\tau)|\beta(t)|^2+C(\tau)|\beta(t)|^4\right)\nonumber\\
  &+& \left(\frac{1}{N}+|S_\infty|^2\right)\left(\frac{2}{N}+|S_\infty|^2\right)\left[\beta(t) B(\tau)+\textrm{c.c.}\right].
\end{eqnarray}
Using the definitions of $A(\tau)$, $B(\tau)$ and $C(\tau)$ and taking the limit $t\rightarrow\infty$, we obtain
\begin{eqnarray}
  g^{(2)}(\tau) &=&
  R+2Q
  \left[e^{-\Gamma_N\tau}-e^{-\Gamma_N\tau/2}\cos(\Delta\tau)\right],
  \label{g2:p}
\end{eqnarray}
where
\begin{equation}
Q=R-\frac{2+N|S_\infty|^2}{1+N|S_\infty|^2}
\end{equation}
We observe that in the  ''chaotic'' limit $N|S_\infty|^2\ll 1$, $R\approx 2$, $Q\ll 1$, and
$g^{(2)}(\tau)\approx 2$.
Conversely, in the ''coherent'' limit $N|S_\infty|^2\gg 1$, $R\approx 1$, $Q\ll 1$, and
$g^{(2)}(\tau)\approx 1$.
More generally, $g^{(2)}(\tau)$ depends on $N$, $\sigma=k_0\sigma_r$, and the detection angle $\theta$. The parameter $Q$ takes its maximum value $Q_{max}=1/2$ for $N|S_\infty|^2=1$, with $R=7/4$. As an example, Fig.\ref{fig3} shows $g^{(2)}(\tau)$ vs $\Gamma_N\tau$ for a spherical Gaussian distribution with $N=10^6$ and $k_0\sigma_r=20$, a laser beam with  $\Delta=5\Gamma_N$, and detection angles $\theta=9^\circ,10^\circ,11^\circ,12^\circ$.
Within a few degrees, the value of $g^{(2)}(\tau)$ changes from $1$ to $2$, with damped oscillations as a function of $\tau$.

\section{Numerical results}\label{numerics}

\begin{figure}[h]
        \centerline{\scalebox{0.5}{\includegraphics{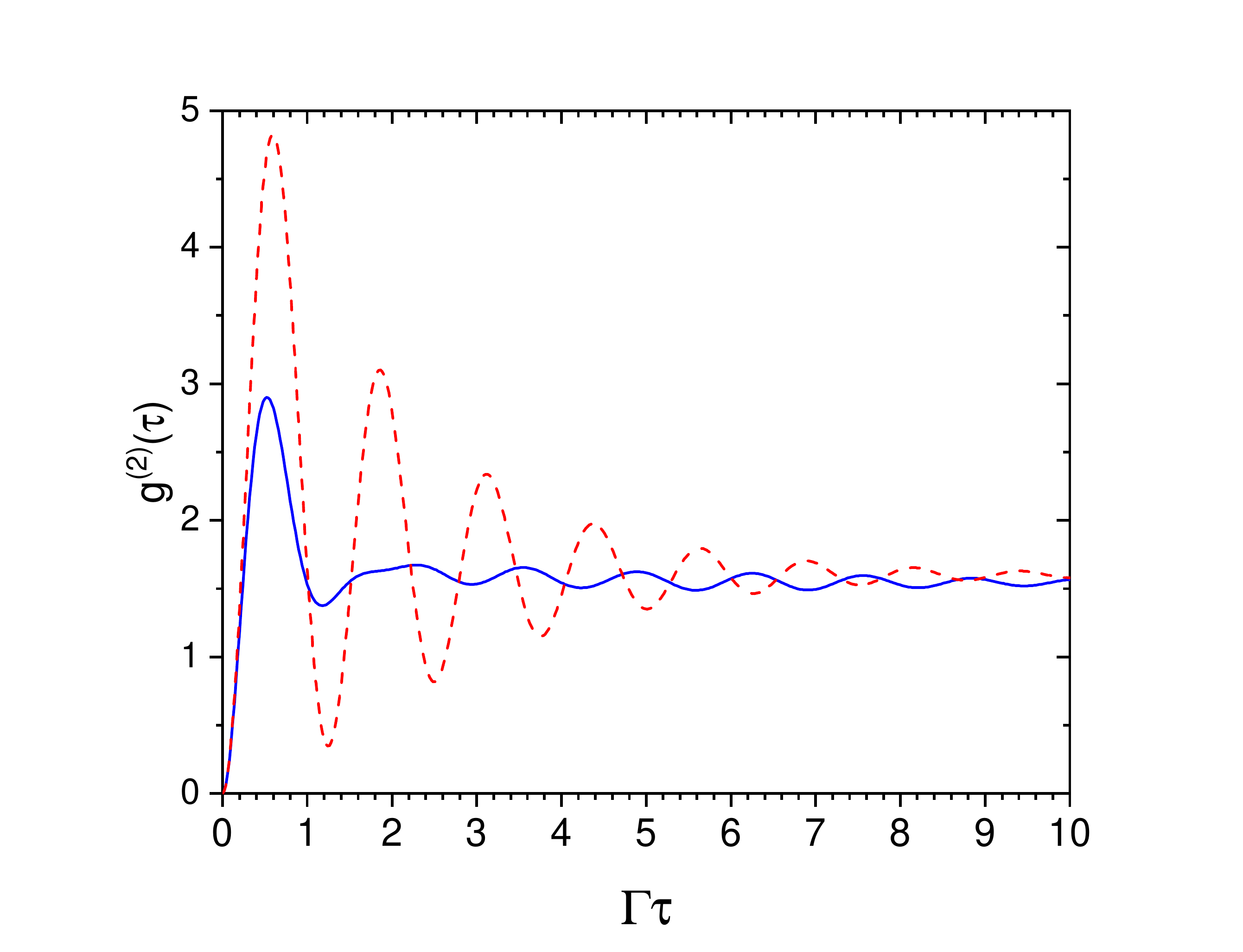}}}
        \caption{$g^{(2)}(\tau)$ vs $\Gamma\tau$, calculated from Eq. (\ref{gg:2:fin}), for $N=100$, $k_0\sigma_r=5$, $\Delta=5\Gamma$, $\theta=90^\circ$ and average after 20 iterations (continuous blue line). The dashed red line is the analytic expression of Eq. (\ref{g2:fin}), with $R=1.7$.}
        \label{fig4}
\end{figure}

\begin{figure}[h]
        \centerline{\scalebox{0.5}{\includegraphics{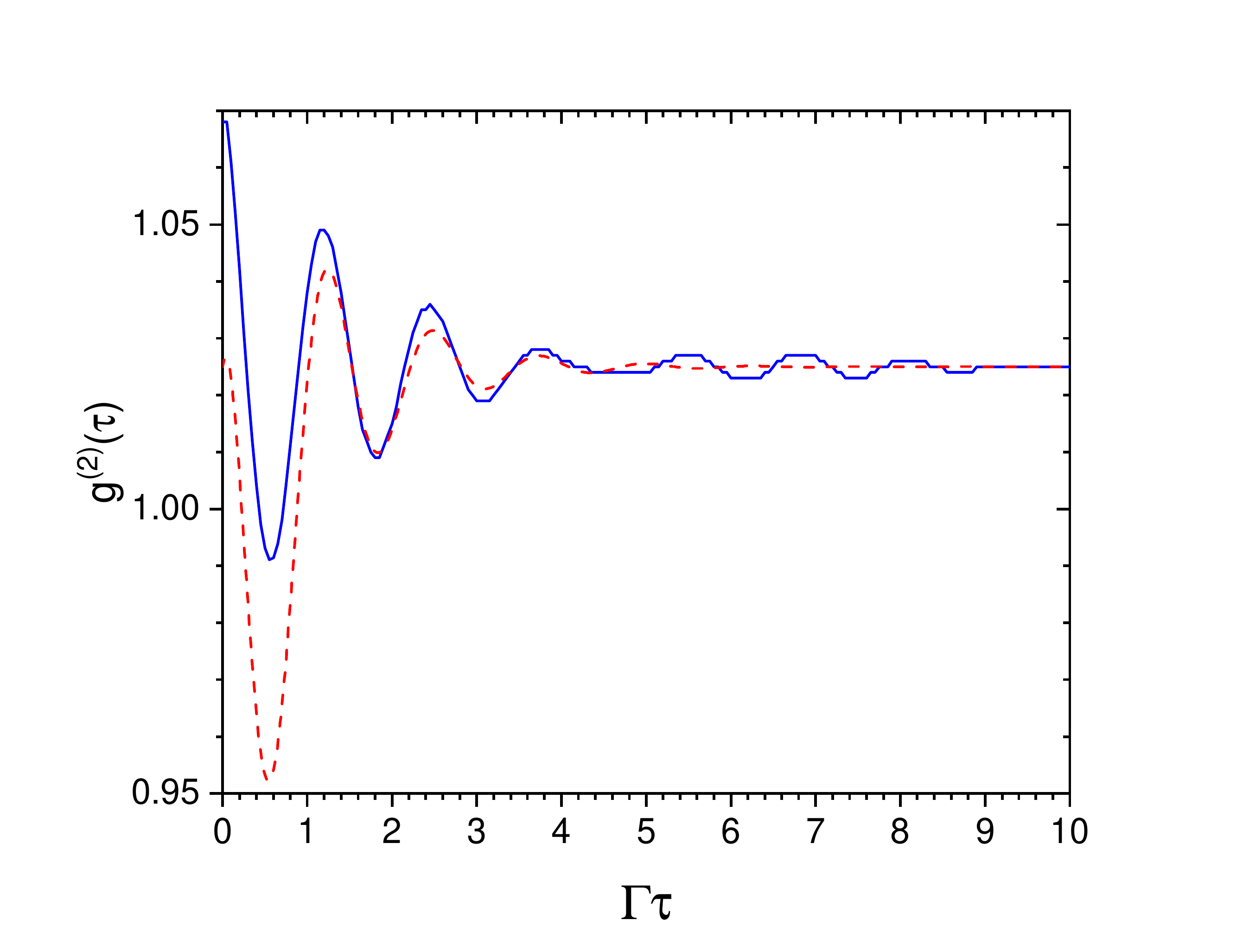}}}
        \caption{$g^{(2)}(\tau)$ vs $\Gamma\tau$, calculated from Eq. (\ref{g2:fin}), for $N=100$, $k_0\sigma_r=5$, $\Delta=5\Gamma$, $\theta=16.26^\circ$ and average after 20 iterations (continuous blue line). The dashed red line is the analytic expression of Eq. (\ref{g2:p}), with R=1.025,$, Q=-0.042$, and $\Gamma_N=2\Gamma$.}
        \label{fig5}
\end{figure}

In this section we compare the exact solution of $g^{(1)}(t,\tau)$ and $g^{(2)}(t,\tau)$, calculated numerically from Eqs. (\ref{gg:1:gen}), (\ref{gg:2:fin}), and (\ref{gg:2:prod}) with the analytic expressions of Eqs. (\ref{g1:beta}), (\ref{g2:beta}), and (\ref{g2:p}). 
The expression of $g^{(2)}(t,\tau)$ assuming $N$ classical dipoles is
\begin{eqnarray}\label{g2:classical}
  g^{(2)}(t,\tau) &=& \frac{1}{I^2(t)}\overline{\sum_{j,m,p,q}
  e^{-i\mathbf{k}\cdot(\mathbf{r}_j-\mathbf{r}_m+\mathbf{r}_q-\mathbf{r}_p)}
    \beta_p^*(t)\beta_m^*(t+\tau)\beta_j(t+\tau)\beta_q(t)}.
\end{eqnarray}
where $I(t)$ is given by Eq.(\ref{I(t)}). In this case $g^{(2)}(\tau)=\lim_{t\rightarrow\infty}g^{(2)}(t,\tau)=R$, i.e. is independent on $\tau$. Hence the oscillations observed in Eq.(\ref{g2:p}) for Eberly's state seem to have a quantum nature (see Fig.\ref{fig3}).
Figure \ref{fig4} shows the result of a numerical simulation calculating $g^{(2)}(t,\tau)$ from Eq.(\ref{gg:2:fin}) for a spherical Gaussian distribution with $N=100$, $k_0\sigma_r=5$, detuning $\Delta=5\Gamma$, detection angle $\theta=90^\circ$, and $\Gamma t=5$, averaged over 20 iterations (solid blue line). The numerical result is compared with the analytic expression of Eq.(\ref{g2:fin}) obtained assuming the timed Dicke state (red dashed line). The numerical value $R\approx 1.7$ obtained for $\Gamma\tau\rightarrow\infty$ is less than the chaotic value 2 because of the small number of particles and iterations. The exact result differs from the approximated timed Dicke solution because of the spread of values of $\beta_j$ around the average value, causing a decoherence which reduces the oscillation amplitudes.
Figure \ref{fig5} shows the result of a numerical simulation calculating $g^{(2)}(t,\tau)$ from Eq.(\ref{gg:2:prod}) for a spherical Gaussian distribution with $N=100$, $k_0\sigma_r=5$, detuning $\Delta=5\Gamma$, detection angle $\theta=16.26^\circ$, and $\Gamma t=5$, averaged over 20 iterations (solid blue line). The numerical result is compared with the analytic expression of Eq. (\ref{g2:p}) obtained assuming Eberly's state with $R\approx 1.025$, $Q\approx -0.042$, and $\Gamma_N\approx 2\Gamma$ (dashed red line). These parameters have been calculated numerically for the spatial distribution. Also in this case, the ideal case of Eberly's state shows oscillations with amplitudes larger than the ones of the exact solution. Nevertheless, the transient oscillations are clearly visible and detectable. We outline again that these oscillations disappear in the classical limit, when the atoms are described as classical dipoles.
\section{Conclusions}

We have calculated the two-time field and intensity correlation functions of the light scattered by an ensemble of two-level atoms driven by a cw resonant laser beam, in the linear optics regime. The atoms have fixed random positions. We have calculated the correlation functions for two different quantum atomic states, i.e., the single-excitation state and the product state, obtaining exact expressions to be evaluated numerically. Furthermore, we have obtained analytic expressions of $g^{(1)}(t,\tau)$ and $g^{(2)}(t,\tau)$ in the case of uniform excitation. This approximation leads to the well-known timed Dicke state \cite{Scully2006} for the single-excitation state and to the pure Bloch state \cite{Friedberg2007} for the product state (named here Eberly's state, in honor of Eberly \cite{Eberly2006} who first discussed the differences between these two states). Our conclusions are that both these states lead, as expected, to $g^{(1)}(\tau)=\lim_{t\rightarrow\infty}g^{(1)}(t,\tau)=1$, as
it results in also describing the atoms as classical dipoles. This result is independent on the statistical properties of the atomic distribution, and corresponds to elastic scattering at the frequency of the incident driving beam. Differences between the single-excitation state and the product state appear when the stationary intensity correlation function is considered, $g^{(2)}(\tau)=\lim_{t\rightarrow\infty}g^{(2)}(t,\tau)$. In the classical limit (i.e., atoms as classical dipoles), $g^{(2)}(\tau)=R$, where $R$ depends on the randomness and spatial distribution of the atomic sample, varying from $R=1$ when the emission is coherent (i.e., within the diffraction cone for an extended cloud) to $R=2$ for a chaotic random distribution. For the timed Dicke state, $g^{(2)}(0)=0$ and $\lim_{\tau\rightarrow\infty}g^{(2)}(\tau)=R$, with a transient time of the order of $1/\Gamma_N$, where $\Gamma_N$ is the cooperative decay rate for $N$ atoms. In this case we observe antibunching, since $g^{(2)}(0)<g^{(2)}(\tau)$. For Eberly's state, $g^{(2)}(0)=R$ and $\lim_{\tau\rightarrow\infty}g^{(2)}(\tau)=R$. However, it is possible to observe transient oscillations as a function of $\tau$ in an intermediate regime with $1<R<2$ and in the detuned case, where the coherent and chaotic emission are competing. These oscillations are purely quantum and are not visible when the atoms are treated as classical dipoles.

We outline again that the aim of this work is to propose a method to distinguish between the possible states generated in the cooperative scattering, by measuring the second-order correlation function  $g^{(2)}(\tau)$. In fact, it gives different results for the single-excitation entangled state or the factorized coherent state. Generally, these states should necessitate a different preparation, as discussed in the original papers by Scully and co-workers \cite{Scully2006,Scully2007} and  more recently studied experimentally  by Felinto and co-workers \cite{Felinto}. Intuitively, it is likely that an ensemble of $N$ two-level atoms driven by a classical field will be described by the product state. However, this has not been proved yet, and an experiment measuring $g^{(2)}(\tau)$ can do it.

This study has assumed a cw driving beam and neglected any atomic motion, either due to temperature or recoil \cite{Robicheaux2021}. It would be interesting in the future to extend it to include the temperature inducing a decay of the correlations (see, for instance, Ref. \cite{Eloy2018}). Also, the statistical properties of subradiance \cite{Guerin2016} should deserve attention, which may be investigated with the present formalism just switching off the driving laser and observing the fluorescence light emitted by the excited atoms at sufficiently long time, such that only the subradiance component survives. All these points will be the object of a future publication.

\acknowledgments

We thank R. Bachelard for useful discussions and R. Gaiba for the analytic calculations of Appendix \ref{Eberly} done during his stage in Milan.
This work was performed within the framework of the European Training Network ColOpt, which is funded by the European Union (EU) Horizon 2020 program under the Marie  Sklodowska-Curie actions, Grant Agreement No. 721465 R.A.
\appendix

\section{Evaluation of $g^{(1)}(\tau)$ and $g^{(2)}(\tau)$ for the single-excitation state (\ref{psi})}\label{g1:g2}

Using Eqs. (\ref{s1:Mark}) with the quantum regression theorem,
\begin{eqnarray}
  \frac{d\langle\sigma_{m}^\dagger(t)\sigma_{j}(t+\tau)\rangle}{d\tau} &=&
  \left(i\Delta_0-\frac{\Gamma}{2}\right)\langle\sigma_{m}^\dagger(t)\sigma_{j}(t+\tau)\rangle-
  \frac{i\Omega_0}{2}e^{i\mathbf{k}_0\cdot\mathbf{r}_j}\beta_m^*(t)-\frac{\Gamma}{2}
  \sum_{k\neq j}\gamma_{jk}
  \langle\sigma_{m}^\dagger(t)\sigma_{k}(t+\tau)\rangle\label{g1mj:A},
\end{eqnarray}
where $\beta_m^*(t)$ is the solution of Eq. (\ref{s1:beta}).  Equation (\ref{g1mj:A}) is integrated with
the initial condition, at $\tau=0$,
$\langle\sigma_{m}^\dagger(t)\sigma_{j}(t)\rangle=\beta_m^*(t)\beta_j(t)$.
Let us note that, setting
$\langle\sigma_{m}^\dagger(t)\sigma_{j}(t+\tau)\rangle=\beta_m^*(t)G_{mj}(t,\tau)$,
Eq. (\ref{g1mj:A}) becomes
\begin{eqnarray}
  \frac{dG_{mj}(t,\tau)}{d\tau} &=&
  \left(i\Delta_0-\frac{\Gamma}{2}\right)G_{mj}(t,\tau)-
  \frac{i\Omega_0}{2}e^{i\mathbf{k}_0\cdot\mathbf{r}_j}-\frac{\Gamma}{2}
  \sum_{k\neq j}\gamma_{jk}G_{mk}(t,\tau)\label{g1mj:bis:A}.
\end{eqnarray}
with $G_{mj}(t,0)=\beta_j(t)$. It is clear from
(\ref{g1mj:bis:A}) that $G_{mj}$ is independent on $m$ and,
comparing Eq.(\ref{g1mj:bis:A}) with Eq.(\ref{s1:beta}),
$G_{mj}(t,\tau)=\beta_j(t+\tau)$, so that
\begin{eqnarray}\label{gg:1:gen:A}
  g^{(1)}(t,\tau) &=& \frac{1}{I(t)}\overline{\sum_{j,m}e^{-i\mathbf{k}\cdot(\mathbf{r}_j-\mathbf{r}_m)}\beta_m^*(t)\beta_j(t+\tau)}e^{-i\omega_0\tau}.
\end{eqnarray}
In order to obtain $g^{(2)}(t,\tau)$, we need to evaluate
$g_{pmjq}(t,\tau)=\langle\sigma_p^\dagger(t)\sigma_m^\dagger(t+\tau)\sigma_j(t+\tau)\sigma_q(t)\rangle$. First,
we obtain the equation
\begin{eqnarray}
  \frac{d}{d\tau}(\sigma^\dagger_{m}\sigma_{j})&=&-\Gamma\sigma^\dagger_{m}\sigma_{j}
  -\frac{i\Omega_0}{2}\left[e^{i\mathbf{k}_0\cdot\mathbf{r}_j}\sigma^\dagger_m-e^{-i\mathbf{k}_0\cdot\mathbf{r}_m}\sigma_j\right]-\frac{\Gamma}{2}
  \left(\sum_{k\neq j}\gamma_{jk}\sigma^\dagger_{m}\sigma_{k}+\sum_{k\neq m}\gamma^*_{mk}\sigma^\dagger_{k}\sigma_{j}\right)\label{ss3:A}.
\end{eqnarray}
and then, from the quantum regression theorem,
\begin{eqnarray}
  \frac{d}{d\tau}g_{pmjq}(t,\tau)&=&-\Gamma g_{pmjq}(t,\tau)-
  \frac{i\Omega_0}{2}\left[e^{i\mathbf{k}_0\cdot\mathbf{r}_j}f^*_{qmp}(t,\tau)-e^{-i\mathbf{k}_0\cdot\mathbf{r}_m}f_{pjq}(t,\tau)\right]
  \nonumber\\
  &-&\frac{\Gamma}{2}
  \left[\sum_{k\neq j}\gamma_{jk}g_{pmkq}(t,\tau)+\sum_{k\neq m}\gamma^*_{mk}g_{pkjq}(t,\tau)\right]\label{ss4:A},
\end{eqnarray}
where
$f_{pjq}(t,\tau)=\langle\sigma_p^\dagger(t)\sigma_j(t+\tau)\sigma_q(t)\rangle$
are solutions of the equations:
\begin{eqnarray}
  \frac{d}{d\tau}f_{pjq}(t,\tau)&=&\left(i\Delta_0-\frac{\Gamma}{2}\right)f_{pjq}(t,\tau)-
  \frac{i\Omega_0}{2}e^{i\mathbf{k}_0\cdot\mathbf{r}_j}\beta^*_p(t)\beta_q(t)-\frac{\Gamma}{2}
  \sum_{k\neq j}\gamma_{jk}f_{pkq}(t,\tau)\label{ss5:A}.
\end{eqnarray}
The initial conditions for Eqs.(\ref{ss4:A}) and (\ref{ss5:A}) are $g_{pmjq}(t,0)=0$ and $f_{pjq}(t,0)=0$.
By setting
$f_{pjq}(t,\tau)=\beta_p^*(t)\beta_q(t)F_{pjq}(t,\tau)$,
Eq.(\ref{ss5:A}) yields
\begin{eqnarray}
  \frac{d}{d\tau}F_{pjq}(t,\tau)&=&\left(i\Delta_0-\frac{\Gamma}{2}\right)F_{pjq}(t,\tau)-
  \frac{i\Omega_0}{2}e^{i\mathbf{k}_0\cdot\mathbf{r}_j}-\frac{\Gamma}{2}
  \sum_{k\neq j}\gamma_{jk}F_{pkq}(t,\tau)\label{ss6:A}.
\end{eqnarray}
with $F_{pjq}(t,0)=0$. Hence $F_{pjq}(t,\tau)$ is independent on $p,q,t$ and, comparing Eq. (\ref{ss6:A}) with Eq. (\ref{s1:beta}),
$F_{pjq}(t,\tau)=\beta_j(\tau)$ and $f_{pjq}(t,\tau)=\beta_p^*(t)\beta_j(\tau)\beta_q(t)$, so that Eq. (\ref{ss4:A}) becomes
\begin{eqnarray}
  \frac{d}{d\tau}g_{pmjq}(t,\tau)&=&-\Gamma g_{pmjq}(t,\tau)-
  \frac{i\Omega_0}{2}\beta_p^*(t)\left[e^{i\mathbf{k}_0\cdot\mathbf{r}_j}\beta^*_m(\tau)
  -e^{-i\mathbf{k}_0\cdot\mathbf{r}_m}\beta_j(\tau)\right]\beta_q(t)
  \nonumber\\
  &-&\frac{\Gamma}{2}
  \left[\sum_{k\neq j}\gamma_{jk}g_{pmkq}(t,\tau)+\sum_{k\neq m}\gamma^*_{mk}g_{pkjq}(t,\tau)\right]\label{gpmjq:A}.
\end{eqnarray}
Setting
$g_{pmjq}(t,\tau)=\beta_p^*(t)H_{pmjq}(t,\tau)\beta_q(t)$,
Eq. (\ref{gpmjq:A}) becomes
\begin{eqnarray}
  \frac{d}{d\tau}H_{pmjq}(t,\tau)&=&-\Gamma H_{pmjq}(t,\tau)-
  \frac{i\Omega_0}{2}\left[e^{i\mathbf{k}_0\cdot\mathbf{r}_j}\beta^*_m(\tau)
  -e^{-i\mathbf{k}_0\cdot\mathbf{r}_m}\beta_j(\tau)\right]
  \nonumber\\
  &-&\frac{\Gamma}{2}
  \left[\sum_{k\neq j}\gamma_{jk}H_{pmkq}(t,\tau)+\sum_{k\neq m}\gamma^*_{mk}H_{pkjq}(t,\tau)\right]\label{Hpmjq:A}
\end{eqnarray}
with $H_{pmjq}(t,0)=0$. It is evident from (\ref{Hpmjq:A}) that
$H_{pmjq}(t,\tau)$ does not depend on $p$, $q$ or
$t$, i.e., $H_{pmjq}(t,\tau)=H_{mj}(\tau)$. Hence the second-order
correlation function is
\begin{eqnarray}\label{gg:2:fin:A}
  g^{(2)}(t,\tau) &=& \frac{1}{I^2(t)}\overline{\sum_{j,m,p,q}e^{-i\mathbf{k}\cdot(\mathbf{r}_j-\mathbf{r}_m+\mathbf{r}_q-\mathbf{r}_p)}
    \beta_p^*(t)H_{mj}(\tau)\beta_q(t)}
\end{eqnarray}
where $H_{mj}(\tau)$ is the solution of the equation
\begin{eqnarray}
  \frac{d}{d\tau}H_{mj}(\tau)&=&-\Gamma H_{mj}(\tau)-
  \frac{i\Omega_0}{2}\left[e^{i\mathbf{k}_0\cdot\mathbf{r}_j}\beta^*_m(\tau)
  -e^{-i\mathbf{k}_0\cdot\mathbf{r}_m}\beta_j(\tau)\right]
  -\frac{\Gamma}{2}
  \left[\sum_{k\neq j}\gamma_{jk}H_{mk}(\tau)+\sum_{k\neq m}\gamma^*_{mk}H_{kj}(\tau)\right]\label{Hmj}.
\end{eqnarray}
with $H_{mj}(0)=0$.

\section{Evaluation of $g^{(2)}(\tau)$ for the product state (\ref{state:product})}\label{product}
In the chain of the derivation of the expectation values
$g_{pmjq}(t,\tau)=\langle\sigma_p^\dagger(t)\sigma_m^\dagger(t+\tau)\sigma_j(t+\tau)\sigma_q(t)\rangle$ with Eq.(\ref{ss4:A}) and 
$f_{pjq}(t,\tau)=\langle\sigma_p^\dagger(t)\sigma_j(t+\tau)\sigma_q(t)\rangle$ with Eq.(\ref{ss5:A}), their initial conditions are
\begin{equation}
g_{pmjq}(t,0)
= \left\{
\begin{array}{ccc} \displaystyle
 \beta^*_p(t)\beta_m^*(t)\beta_j(t)\beta_q(t) & & \mathrm{if}\quad p\neq m\quad \mathrm{or}\quad  j\neq q\\
 & & \\
 0 &   & \mathrm{otherwise}
\end{array}\right..
\end{equation}
and
\begin{equation}
f_{pjq}(t,0)
= \left\{
\begin{array}{ccc} \displaystyle
 \beta_p^*(t)\beta_j(t)\beta_q(t) & \mathrm{if} & j\neq q \\
 & & \\
 0 &  & \mathrm{otherwise}
\end{array}\right..
\end{equation}
By setting
$f_{pjq}(t,\tau)=\beta_p^*(t)\beta_q(t)F_{pjq}(t,\tau)$,
Eq.(\ref{ss5:A}) yields
\begin{eqnarray}
  \frac{d}{d\tau}F_{pjq}(t,\tau)&=&\left(i\Delta_0-\frac{\Gamma}{2}\right)F_{pjq}(t,\tau)-
  \frac{i\Omega_0}{2}e^{i\mathbf{k}_0\cdot\mathbf{r}_j}-\frac{\Gamma}{2}
  \sum_{k\neq j}\gamma_{jk}F_{pkq}(t,\tau)\label{Fpjq:B}.
\end{eqnarray}
with 
\begin{equation}
F_{pjq}(t,0)
= \left\{
\begin{array}{ccc} \displaystyle
\beta_j(t) & \mathrm{if} & j\neq q \\
 & & \\
 0 &  & \mathrm{otherwise}
\end{array}\right..
\end{equation}
Hence $F_{pjq}(t,\tau)$ is independent on $p,q$, $F_{pjq}(t,\tau)=F_j(t,\tau)$. Comparing Eq. (\ref{Fpjq:B}) with Eq. (\ref{s1:beta}), we deduce that
$F_{j}(t,\tau)=\beta_j(t+\tau)$ if $j\neq q$ and $F_{j}(\tau)=\beta_j(\tau)$ if $j=q$. The same reasoning will be true for $g_{pmjq}(t,\tau)$: from
 Eq.(\ref{ss4:A}),
\begin{eqnarray}
  \frac{d}{d\tau}g_{pmjq}(t,\tau)&=&-\Gamma g_{pmjq}(t,\tau)-
  \frac{i\Omega_0}{2}\beta_p^*(t)\left[e^{i\mathbf{k}_0\cdot\mathbf{r}_j}F^*_m(t,\tau)
  -e^{-i\mathbf{k}_0\cdot\mathbf{r}_m}F_j(t,\tau)\right]\beta_q(t)
  \nonumber\\
  &-&\frac{\Gamma}{2}
  \left[\sum_{k\neq j}\gamma_{jk}g_{pmkq}(t,\tau)+\sum_{k\neq m}\gamma^*_{mk}g_{pkjq}(t,\tau)\right]\label{gpmjq:2:B}.
\end{eqnarray}
Setting
$g_{pmjq}(t,\tau)=\beta_p^*(t)H_{pmjq}(t,\tau)\beta_q(t)$,
Eq. (\ref{gpmjq:2:B}) becomes
\begin{eqnarray}
  \frac{d}{d\tau}H_{pmjq}(t,\tau)&=&-\Gamma H_{pmjq}(t,\tau)-
  \frac{i\Omega_0}{2}\left[e^{i\mathbf{k}_0\cdot\mathbf{r}_j}F^*_m(t,\tau)
  -e^{-i\mathbf{k}_0\cdot\mathbf{r}_m}F_j(t,\tau)\right]
  \nonumber\\
  &-&\frac{\Gamma}{2}
  \left[\sum_{k\neq j}\gamma_{jk}H_{pmkq}(t,\tau)+\sum_{k\neq m}\gamma^*_{mk}H_{pkjq}(t,\tau)\right]\label{Hpmjq:2:B}
\end{eqnarray}
with
\begin{equation}
H_{mj}(t,0)
= \left\{
\begin{array}{ccc} \displaystyle
 \beta_m^*(t)\beta_j(t) & & \mathrm{if}\quad p\neq m\quad \mathrm{or}\quad  j\neq q\\
 & & \\
 0 &   & \mathrm{otherwise}
\end{array}\right..
\end{equation}
It is evident from (\ref{Hpmjq:2:B}) that
$H_{pmjq}(t,\tau)$ depends on $p,q$ only for the initial condition at $\tau=0$, so that $H_{pmjq}(t,\tau)=H_{mj}(t,\tau)$. 
Hence, the second-order correlation function is
\begin{eqnarray}\label{gg:2:prod:B}
  g^{(2)}(t,\tau) &=& \frac{1}{I^2(t)}\overline{\sum_{j,m,p,q}e^{-i\mathbf{k}\cdot(\mathbf{r}_j-\mathbf{r}_m+\mathbf{r}_q-\mathbf{r}_p)}
    \beta_p^*(t)H_{mj}(t,\tau)\beta_q(t)}
\end{eqnarray}
where $H_{mj}(\tau)$ is the solution of the equation
\begin{eqnarray}
  \frac{d}{d\tau}H_{mj}(t,\tau)&=&-\Gamma H_{mj}(t,\tau)-
  \frac{i\Omega_0}{2}\left[e^{i\mathbf{k}_0\cdot\mathbf{r}_j}F^*_m(t,\tau)
  -e^{-i\mathbf{k}_0\cdot\mathbf{r}_m}F_j(t,\tau)\right]\nonumber\\
  &-&\frac{\Gamma}{2}
  \left[\sum_{k\neq j}\gamma_{jk}H_{mk}(t,\tau)+\sum_{k\neq m}\gamma^*_{mk}H_{kj}(t,\tau)\right]\label{Hmj:prod:B}.
\end{eqnarray}

\section{Evaluation of $\langle E^\dagger(t)E(t) E(t)\rangle $ and $\langle E^\dagger(t)E^\dagger(t)E(t) E(t)\rangle$ for the Eberly's state (\ref{state:Eberly})} \label{Eberly}

 From the definition, we have
\begin{eqnarray}
  \langle E^\dagger(t)E(t) E(t)\rangle &=& \frac{1}{N^3}\sum_{jmp} e^{i(\mathbf{k}_0-
    \mathbf{k})\cdot(\mathbf{r}_m+\mathbf{r}_p-\mathbf{r}_j)}\langle\sigma^\dagger_j\sigma_m \sigma_p\rangle\\
  \langle E^\dagger(t)E^\dagger(t)E(t) E(t)\rangle &=& \frac{1}{N^4}\sum_{jmpq} e^{i(\mathbf{k}_0-
    \mathbf{k})\cdot(\mathbf{r}_p+\mathbf{r}_q-\mathbf{r}_j-\mathbf{r}_m)}\langle\sigma^\dagger_j\sigma^\dagger_m\sigma_p \sigma_q\rangle
\end{eqnarray}
and the not vanishing expectation values are
\begin{equation}\label{sss}
\langle\sigma^\dagger_j\sigma_m \sigma_p\rangle= \beta|\beta|^2
\left\{
\begin{array}{cc} \displaystyle
 \mathrm{if} & m=j\neq p\\
 & \\
 \mathrm{if} & p=j\neq m\\
 & \\
 \mathrm{if} & p\neq j\neq m
\end{array}
 \right.
\end{equation}
and
\begin{equation}\label{ssss}
\langle\sigma^\dagger_j\sigma^\dagger_m\sigma_p \sigma_q\rangle=
|\beta|^4 \left\{
\begin{array}{cc} \displaystyle
 \mathrm{if} & j=p\neq m=q\\
 & \\
 \mathrm{if} & j=q\neq m=p\\
 & \\
 \mathrm{if} & j=p,j\neq m,j\neq q,m\neq q,\\
 & \\
 \mathrm{if} & j=q,j\neq m,j\neq p,p\neq m,\\
 & \\
 \mathrm{if} & m=p,m\neq q,m\neq j,q\neq j,\\
 & \\
 \mathrm{if} & m=q,p\neq m,p\neq j,m\neq j,\\
 & \\
 \mathrm{if} & p\neq j\neq m\neq q
\end{array}
 \right.
\end{equation}
From them we obtain:
\begin{eqnarray}
  \langle E^\dagger(t)E(t) E(t)\rangle &=&
    \beta|\beta|^2\left\{
    \frac{2(N-1)}{N^2}+K_2\right\}\label{EEE}\\
  \langle E^\dagger(t)E^\dagger(t)E(t) E(t)\rangle &=& |\beta|^4\left\{
    \frac{2(N-1)}{N^3}+\frac{4(N-2)}{N^2}K_2+K_4\right\}\label{EEEEbis}
\end{eqnarray}
where
\begin{eqnarray}
  K_2 &=& \frac{1}{N^2}\sum_{j}\sum_{p\neq j} e^{i(\mathbf{k}_0-
    \mathbf{k})\cdot(\mathbf{r}_p-\mathbf{r}_j)}\approx|S_\infty|^2\\
     K_4 &=&  \frac{1}{N^4}\sum_{m}\sum_{j\neq m}\sum_{p\neq j,m} \sum_{q\neq p,j,m}
    e^{i(\mathbf{k}_0-
    \mathbf{k})\cdot(\mathbf{r}_p+\mathbf{r}_q-\mathbf{r}_j-\mathbf{r}_m)}\approx|S_\infty|^4.
\end{eqnarray}

\end{document}